\documentclass[epj-spec]{svjour}
\usepackage{graphics}
\usepackage{amsmath,bm,epsfig}
\usepackage{epsfig,color}
\usepackage{bm} 
\usepackage{ulem}
\newcommand{\ee}{\end{equation}}
\newcommand{\be}{\begin{equation}}
\newcommand{\bea}{\begin{eqnarray}}
\newcommand{\eea}{\end{eqnarray}}

\renewcommand{\it}[1]{\textit{#1}}

\begin{document}

\title{Population dynamics in compressible flows}
\author{ Roberto Benzi$^{(1)}$, Mogens H. Jensen$^{(2)}$,David R. Nelson$^{(3)}$  , Prasad Perlekar $^{(4)}$,  Simone Pigolotti $^{(5)}$ , Federico Toschi $^{(4)}$}  

\institute{ $^{(1)}$ Dip. di Fisica and INFN, Universit\`a ``Tor
  Vergata", Via della Ricerca Scientifica 1, I-00133 Roma, Italy. \\
  $^{(2)}$The Niels
  Bohr Institut, Blegdamsvej 17, DK-2100 Copenhagen, Denmark. \\
  $^{(3)}$ Lyman Laboratory of Physics, Harvard University, Cambridge, MA 02138, USA\\
  $^{(4)}$ Department of Physics, Department of Mathematics and
  Computer Science, and J.M. Burgerscentrum, Eindhoven University of
  Technology, 5600 MB Eindhoven, The Netherlands; \\and
  International Collaboration for Turbulence Research. \\
  $^{(5)}$ Dept. de Fisica i Eng. Nuclear, Universitat Politecnica de
  Catalunya Edif. GAIA, Rambla Sant Nebridi s/n, 08222 Terrassa,
  Barcelona, Spain }

 
\abstract{Organisms often grow, migrate and compete in liquid environments,
as well as on solid surfaces. However, relatively little is known about
what happens when competing species are mixed and compressed
by fluid turbulence. In these lectures  we review our recent work on population dynamics and population genetics in 
compressible velocity fields of one and two dimensions. We discuss why compressible turbulence is relevant for population dynamics
in the ocean and  we
consider cases both where the velocity field is turbulent and 
when it is static. Furthermore, we investigate populations 
in terms of a continuos density field and when the populations are treated via
discrete particles. In the last case we focus on
the competition and fixation of one species compared
to another}

\maketitle

\section{Introduction} 

Challenging problems arise when spatial migrations of species are
combined with population genetics.  Stochastic number fluctuations are
inevitable at a frontier, where the population size is small and the
discrete nature of the organisms becomes essential. Depending on the
parameter values, these fluctuations can produce important changes
with respect to the deterministic predictions \cite{3,4}.  When two or
more species undergo a Darwinian competition in a spatial
environment, one must deal with additional issues such as genetic
drift (stochastic fluctuations in the local fraction of one species
compared to another) and Fisher genetic waves, \cite{fisher} which
allow more fit species to replace less fit ones. On solid surfaces,
the complexities of spatial population genetics are elegantly
accounted for by the stepping stone model, originally introduced by
Kimura and Weiss \cite{6}, \cite{8}.

However, much of population genetics, from the distant
past up to the present, played out in liquid environments,
such as lakes, rivers and oceans. For example, there are fossil
evidence for oceanic photosynthetic cyanobacteria (likely pre-
cursors of chloroplasts in plants and a major source of oxygen
in the atmosphere) that date back a billion years or more \cite{11}.
In addition, it has recently become possible to perform satellite 
observations of chlorophyll concentrations to identify fluid
dynamical niches of phytoplankton types off the eastern coast
of the southern tip of South America \cite{12}, where the domains
of the species  are largely determined by the tangen-
tial velocity field obtained from satellite altimetry. In cases
such as these, spatial growth and evolutionary competition
take place in the presence of advecting flows, some of them at
high Reynolds numbers \cite{13}. 

{ Phytoplankton needs light and nutrients to grow and many
  phytoplankton species are able to adjust their density and swim to
  stay near the surface. Nutrients are brought to the surface from
  deeper ocean layers, usually below 500 meters. Therefore, oceanic
  circulation plays an important role in shaping spatial growth and
  evolution of plankton species.  To appreciate the complexity of the
  problem, it is worthwhile to shortly review our present knowledge on
  the basic mechanisms one should consider.  In a three dimensional
  turbulent flow at high Reynolds number, the velocity field is
  fluctuating over a range of scales $[L, \eta]$ where $L$ is the
  scale of energy pumping in the system and $\eta \equiv
  (\nu^3/\epsilon)^{1/4}$ is the Kolmogorov dissipation scale. The
  velocity field is also fluctuating in time. According to Kolmogorov
  theory, one can define the dissipation time scale as $\tau_{\eta}
  \equiv \sqrt{\nu/\epsilon}$. In the upper oceanic mixed layer ,
  forcing is provided by heat and momentum exchange with atmosphere
  and the observed values \cite{dissipation} of $\epsilon$ ranges from
  $10^{-7} cm^2/sec^3$ up to $50 cm^2/sec^3$, which implies $\eta \in
  [0.01 , 2] cm$ and $\tau_{\eta} \in [0.01, 300] sec$. The
  phytoplankton size lies in the range $[10 ,200 ] \mu m$ with a
  density difference respect to sea water density in the range $[0.01
  , 0.1]$.  Advection of individuals in the ocean should be studied by
  considering all forces acting on them. In particular, because of
  density mismatch and finite size, individuals are not advected as
  simple Lagrangian tracers \cite{Toschi} \cite{14} , i.e. the
  velocity field experienced by each individual is not the Lagrangian
  velocity field, but an effective velocity field which may be not
  incompressible. A suitable measure of compressibility can be defined
  as $ \kappa = \langle (div\ \vec{v})^2 \rangle / \langle
  (\vec{\nabla} \vec{v})^2 \rangle $, where $\langle .. \rangle$
  stands for space and time average. Using the above mentioned values
  of phytoplankton size $a$, density mismatch $\delta \rho / \rho$ and
  turbulent energy dissipation $\epsilon$, one obtains
$$
\sqrt{\kappa} = \frac{\delta \rho}{\rho} \frac{a^2}{\nu \tau_{\eta}}  \in [10^{-9}, .4] 
$$
Another very important feature to be considered is the ability of
individuals to swim in a preferential direction towards the largest
concentration of nutrients (chemotaxis). The swimming velocity $ V_c$
is presently estimated in the range $[10 , 500 ] \mu m /sec$. Because
of turbulent, individuals are subject to external forces which try to
change the direction. It is observed that with a characteristic time
$B \sim 5 sec$, individuals try to recover the preferential
direction. This mechanism, named gyrotaxis \cite{Gyrotaxis1} and
\cite{Gyrotaxis2}, introduces an effective compressible flow with
compressibility
$$
\sqrt{\kappa} = \frac{V_c B }{ \eta} \in [2.5 10^{-3}, 1]
$$
It is important to remark that turbulent flows with an effective
compressibility can dramatically change population dynamics:
concentration of individuals increases in  low pressure regions
(sinks) and decreases in high pressure regions (source) and the
population is spatially characterized by small scale patchiness. The
above discussion shows that intense turbulent activity in the oceanic
upper layer may introduce non trivial effect, due to compressibility,
in the phytoplankton growth and evolution at rather small scale.  }
\bigskip { The same considerations might be relevant for large scale
  motions. Very large scale oceanic circulation ($100 - 300 km$) are
  characterized by relatively small Rossby number $Ro$, defined as $Ro
  = u_H / (fL)$, where $u_H$ is the characteristic horizontal
  velocity, order $0.1 m/sec$, $f=10^{-4} sec^{-1}$ is the Coriolis
  frequency and $L$ is the characteristic large scale circulation. For
  $Ro<<1$, the velocity field is close to the geostrophic balance,
  meaning that the Coriolis force balances the pressure gradient.
  Under such circumstances, the vertical velocity $w$ is rather small
  and it can be estimated to be $0.1 mm/sec$ or equivalently few
  meters/day. The horizontal velocity can be decomposed in the
  geostrophic component $\vec{v}_g$ and the non geostrophic part
  $\vec{v}_a$ where $div_H \vec{v}_g= 0$ and $div_H \vec{v}_a
  + \partial_z w = 0$ with $div_H \equiv \partial_x + \partial_y$.
  According to quasi-geostrophic dynamics, near the surface there
  exists an effective compressible flow acting on time scale order
  $div\ \vec{v}_a \sim 10^{-6} sec^{-1}$ much longer than the longest
  population growth rate $\mu \sim 2\ 10^{-5} sec^{-1}$. Therefore, at
  very large scale, population dynamics evolves under the advection of
  an incompressible flow. The above picture changes dramatically if we
  consider flows at $Ro$ close to $1$. Recent numerical simulations as
  well as direct observations \cite{klein},\cite{Mizobota}, \cite{15}
  have shown that surface density tends to develop sharp horizontal
  gradients (fronts) especially near by the edge of oceanic
  eddies. Formation of intense fronts, produced by the enhanced
  filamentation of surface density \cite{SQG1} \cite{SQG2}, increases
  the vertical advection and destroy geostrophic balance.  As a
  results two important phenomena seem to take place in the ocean at
  relatively large scale (order $10 km$) \cite{3docean}\cite{fronts1}:
  regions of relative large and positive vertical velocity (upwelling)
  tends to increase nutrients for phytoplankton providing an increase
  of total biological mass while regions of negative vertical velocity
  increases concentration of the phytoplankton
  population. Frontogenesis, as it is usually named the formation of
  sharp density gradients, can develop vertical velocity up to few
  millimeter/sec. Consequently, the horizontal velocity near
  frontogenetic regions is characterized by an effective
  compressibility with $div_H \vec{v} \sim 10^{-4}$ \cite{3docean},
  i.e. smaller than the population growing rate. The above picture
  suggests the formation of plankton patchiness on scale ranging from
  $100 m$ to $10-30 km$.  } { As a tentative conclusion to our short
  review of phytoplankton in the ocean, albeit the complexity of the
  problem, it seems important to understand the role of turbulent
  compressible flows in population dynamics and population genetics
  trying to understand, at least in the simplest cases, if a new and
  non trivial phenomenology can be discovered and its relevance to
  biological evolution.  }


{In these lectures  we review our recent work \cite{17},\cite{18},\cite{18b} on population dynamics and population genetics in 
compressible velocity fields of one and two dimensions, motivated by the above discussion.  We
consider cases both where the velocity field is turbulent and 
when it is static. Furthermore, we investigate populations 
in terms of a continuos density field and when the populations are treated via
discrete particles. In the last case we focus on
the competition and fixation of one species compared
to another. }

\section{One dimensional case}

In this section we shall discuss some qualitative and quantitative ideas underlying the effect of compressible turbulence on population dynamics. We restrict ourself to the one dimensional case where most concepts
can be discussed using rather simple analytical tools.

Upon specializing to one dimension, the Fisher equation reads \cite{2}
\be
\label{1d}
\partial_t c + \partial_x (u c)  = D \partial_{x}^2 c + \mu c- b c^2
\ee
Equation (\ref{1d}) is relevant for the case of compressible flows, 
where $\partial_x u \ne 0$, and for the case
when the field $c(x,t)$ describes the population of 
inertial particles or biological species.
By suitable rescaling of $c(x.t)$, we can always set $b=\mu$. In the following, unless stated otherwise, 
we shall assume $b=\mu$ whenever $\mu \ne 0$ and $b=0$ for $\mu=0$. 

\bigskip

\begin{figure}[h]
  \begin{center}
    \includegraphics[width=0.5\textwidth]{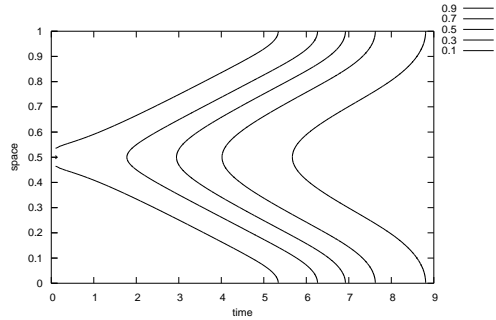}
  \end{center}
  \caption{ Contour plot of the numerical simulation of eq. \ref{1d} with $\mu=1$, $D=0.005$ and with periodic boundary conditions. The initial conditions are $c(x,t)=0$ everywhere expect for few grid points near $L/2=0.5$ where $c=1$. The horizontal axis represents time while the vertical axis is space. }
  \label{figure1}
\end{figure}

The Fisher equation for $u=0$ has travelling
front solutions which can be computed analytically:
\begin{equation}
\label{solution}
c(x,t) = \frac{1}{ [ 1+ C exp(-5 \mu t/6 \pm x \sqrt{\mu/D} /6)]^2}
\end{equation}
From (\ref{solution}) we can see that the non linear wave propagates 
with velocity $v_F \sim (D\mu)^{1/2}$ \cite{fisher}, \cite{2}.
In Fig. (\ref{figure1}) we show a numerical solution of Eq.  (\ref{1d}) with $D=0.005$, $\mu= 1$ and $u=0$ obtained by numerical
integration on a space domain of size $L=1$ with periodic boundary conditions. The figure shows the space-time behaviour 
of $c(x,t)$ for  $c(x,t) = 0.1,0.3.0,5.0,7$ and $0.9$. With initial condition $c(x,t=0)$ nonzero  on only a few grid points centered at $x=L/2$, $c(x,t)$
spreads with a velocity $v_F \sim 0.07 $ and, after a time $L/v_F \sim 4$ reaches the boundary.  Note that the characteristic size of the Fisher'wave interface thichness is order $\sqrt{D/\mu}$.

Let us consider first
the case {$\mu=b=0$}. In this limit, Eq. (\ref{1d}) is just the Fokker-Planck
equation describing the probability distribution $P(x,t) \equiv c(x,t) $ to find a particle in the range $(x, x+dx)$ at time $t$, whose
dynamics is given by the stochastic differential equation:
\be
\label{particle}
\frac{dx}{dt} = u(x,t) + \sqrt{ 2D} \eta(t) 
\ee
where $\eta(t)$ is a white noise with $\langle \eta(t)\eta(t')\rangle = \delta(t-t')$. Let us assume for the moment that $u(x,t)=u(x)$ is time independent and, moreover, let us take $u(x,t) = -\Gamma (x-x_0)$. 
Then,
the stationary solution of (\ref{1d}) is given by 
\be
\label{7}
P(x,t) = A^{-1} exp[-\Gamma(x-x_0)^2/2D]
\ee
where $A$ is a normalization constant. $ P(x,t)=P(x)$ is strongly peaked near the points $x_0$ and (\ref{7}) tells us that $P$  spreads around $x_0$ with a characteristic length  
of order $\xi \equiv \sqrt{D/\Gamma_0}$. Hereafter, we shall refer to $\xi$ as "quasi-localization length". 

The same argument can be used to study the  effect for a more generic turbulent like one dimensional field $u(x)$, still time independent.
We can identify  $\Gamma$ as a typical  gradient of the 
turbulent velocity field $u$. In a turbulent flow, the velocity field is correlated over spatial scale
of order $v_*/\Gamma$ where $v_*^2/2$ is the average kinetic energy of the flow. For $P$ to
be localized near a generic  sink at the point $x_0$, despite spatial variation in the turbulent field,  we must require that the localization length $\xi$ should be smaller
than the turbulent correlation scale $v_*/\Gamma$, i.e.
\be
\sqrt{\frac{D}{\Gamma}} < \frac{v_*}{\Gamma} \rightarrow \frac{v_*^2}{D\Gamma} > 2
\label{prima}
\ee
Condition (\ref{prima}) can be easily understood by considering the simple case of a periodic velocity field
$u$, i.e. $u = v_* sin(x v_* /\Gamma)$. In this case, condition (\ref{prima}) states that
$D$ should be small enough for the probability $P$ not to spread over all the minima of $u$.
For small $D$ or equivalently for large $v_*^2/\Gamma$, 
the solution will be localized near the minima
of $u$, at least for the case of a frozen turbulent velocity field $u(x)$.

\bigskip

The above analysis can be extended for velocity field $u(x,t)$ that depend on both space and time. 
The crucial observation is
that,  close to the sinks  $x_i$ of $u(x,t)$, we should have 
$u(x_i,t) \sim 0$. 
Thus, 
although $u$ is a time dependent function, sharp peaks in 
 $P(x,t)$ move quite slowly, simply because $u(x,t) \sim 0$ near the maximum of $P(x,t)$. 
 One can consider
a Lagrangian path $x(t)$ such that $x(0)=x_0$, where $x_0$ is one particular point 
where $u(x_0,0)=0$ and $\partial_x u(x,0)|_{x=x_0} < 0$.
From direct numerical simulation of Lagrangian particles in fully
developed turbulence, we know that the acceleration of Lagrangian particles is a  strongly 
intermittent quantitiy, i.e. it is small most of the time with large (intermittent) bursts. Thus,
we expect that the localized solution of $P$ follows $x(t)$ for quite long times except for
intermittent bursts in the turbulent flow. During such bursts, the position where
$u=0$ changes abruptly, i.e. almost discontinuosly from one point, say $x(t)$, to
another point $x(t+\delta t)$. During the short time interval $\delta t$, $P$ will drift and spread,
eventually reforming to become  localized again near $x(t+\delta t)$.
The above discussion suggests that the probability $P(x,t)$ will be
localized most of the  time in the Lagrangian frame, except for
short time intervals $\delta t$ during an intermittent burst.

\bigskip

From (\ref{prima})  we conclude that for large value of $D$
$P(x,t)$ is spread out,  while for small enough $D$, $P$ should be a localized or sharply peaked function
of $x$ most of the time. An abrupt transition, or at least a 
sharp crossover, from extended to sharply peaked functions  $P$, should
be observed for decreasing $D$.

\bigskip 

It is relatively simple to extend the above analysis for a non zero growth rate $\mu > 0$, see also \cite{vulpiani2} for a time independent flow. 
The requirement (\ref{prima}) is now only
a necessary condition to observe localization in $c$. For $\mu>0$ we must also require
that the characteristic gradient on scale $\xi$ must be larger than $\mu$, i.e. the effect of the {small scale turbulent fluctuations}
should act on a time scale smaller than $1/\mu$. We  estimate the gradient on scale
$\xi$ as $\delta v(\xi)/\xi$, where $\delta v(\xi)$ is
the characteristic velocity difference on scale $\xi$. We invoke 
the Kolmogorov theory, and set
$\delta v(\xi) = v_* (\xi/L)^{1/3}$ to obtain:
\be
\mu < \frac{\delta v(\xi)}{\xi} = \frac{v_* \xi^{-2/3}}{L^{1/3}} = v_*(\frac{\Gamma}{LD})^{1/3}
\label{seconda}
\ee
In (\ref{seconda}), we interpret $\Gamma $ as the characteristic velocity gradient of the turbulent
flow.   Note also that $\delta v(\xi)/\xi \le \Gamma$ 
on the average, which leads to the inequality:
\be
\mu < \Gamma
\label{due}
\ee
From (\ref{prima}) and
(\ref{due}) we also find
\be
\frac{v_*^2}{D \mu} > 2
\ee
a second necessary condition.

One may wonder whether a non zero growth rate  $\mu$ can change our previous conclusions about
the temporal
behavior, and in particular about its effect on the dynamics of the Lagrangian points where
$u(x,t)=0$. Consider the solution of (\ref{1d}) at time $t$, allow for a spatial domain of size $L$, and introduce
the average position
\be
\label{xm}
x_m \equiv \int_0^L dx x \frac{c(x,t)}{Z(t)}
\ee
where $Z(t)=\int_0^L dx c(x,t)$.
Upon assuming for simplicity a single localized solution,
 we can think of $x_m$ just as the position where most of the bacterial concentration $c(x,t)$ is localized.  
 We can compute the time derivative $v_m(t) = dx_m/dt$. After a short computation, we obtain:
\be
\label{vm}
v_m(t) = Z \int_0^L dx (x_m-x) P(x,t)^2 + \int_0^L u(x,t) P(x,t) dx
\ee
where $P(x,t) \equiv c(x,t)/Z(t)$. Note that $v_m$ is
independent of $\mu$. Moreover,
 when $c$ is localized near
$x_m$, {\it both} terms on the r.h.s. of (\ref{vm}) are close to zero. Thus, 
$v_m$ can be significantly different from zero only if
$c$ is no longer localized and the first integral on the r.h.s becomes relevant.
We can now understand
the effect
of the non linear term in (\ref{1d}): when $c(x,t)$ is localized, the non linear term { does not affect the value of $v_m$} 
simply because $v_m$ is close to $0$. On the other hand, when
$c(x,t)$ is extended the non linear term drives the system to  the state $c=1$ which is an exact solution in the 
absence of turbulent convection
 $u(x,t)=0$. 

\bigskip

We now discuss whether our previous analysis can be compared against numerical simulations of (\ref{1d})  in the one dimensional case.
To completely specify equation (\ref{1d}) we must define the dynamics of the "turbulent" velocity field
$u(x,t)$. 
Although we consider a one dimensional case, we want to study the
statistical properties of $c(x,t)$ subjected to turbulent fluctuations which are close
to those generated by the three dimensional 
Navier-Stokes equations. 
Hence, the statistical properties of $u(x,t)$ should be
characterized by intermittency both in space and in time. {Although intermittency is not a crucial point in our
investigations, we want to use a one dimensional velocity field with some generic features in terms of space and time dynamics.
For this reason, }
we
build the turbulent field $u(x,t)$  by appealing to a simplified  shell model of fluid
turbulence \cite{biferale}. The
wavenumber space is divided into shells of scale $k_n = 2^{n-1} k_0$, $n=1,2,...$. For each shell with
characteristic wavenumber $k_n$,  we describe 
turbulence by using the complex Fourier-like variable $u_n(t)$, satisfing the following equation of motion:
\bea
  (\frac{d }{dt}&+&\nu k_n^2 ) u_n = i(k_{n+1} u_{n+1}^* u_{n+2}-\delta
  k_n u_{n-1}^* u_{n+1} \nonumber  \\
&+&(1-\delta) k_{n-1} u_{n-1} u_{n-2})+f_n \ . 
\label{sabra}
\eea
The model contains one free parameter, $\delta$,
and it conserves two quadratic invariants (when the force and the
dissipation terms are absent) for all values of $\delta$. The first is
the total energy $\sum_n |u_n|^2$ and the second is $\sum_n (-1)^n
k_n^{\alpha} |u_n|^2$, where $\alpha= \log_{2} (1-\delta)$.  
In this
note we fix $\delta = -0.4$. For this value of $\delta$ the model
reproduces intermittency features of the real three dimensional Navier Stokes equation with surprising
good accuracy \cite{biferale}. Using $u_n$, we can build the real one dimensional velocity field $u(x,t)$ as follows:
\be
u(x,t) = F\sum_n [ u_n e^{i k_n x} + u^*_n e^{-i k_n x} ],
\ee
where $F$ is a free parameter to tune the strength of velocity fluctuations
(given by $u_n$) relative to other parameters in the model (see next section).
In all numerical simulations we use a forcing function  $f_n = (\epsilon(1+i)/u^{*}_1 )\delta_{n,1}$, i.e. energy is supplied only
to the largest scale corresponding to $n=1$.
With
this choice, the input power in the shell model is simply given by $1/2\sum_n [u^{*}_n f_n + u_n f^{*}_n] = \epsilon$ , i.e. 
it is constant in time.  
To solve Eqs. (\ref{1d}) and (\ref{sabra})  we use a finite difference scheme with periodic boundary conditions.
Theses model equations can be studied in detail without
major computational efforts. 
The free parameters of the model are the diffusion constant $D$, the size of the periodic 1d spatial domain $L$,
the growth rate $\mu$, the viscosity $\nu$ (which fixes the Reynolds number
 $Re$), 
the ``strength' of the turbulence $F$ and finally the power input in the shell model, namely 
$\epsilon$. Note that according to the Kolmogorov theory \cite{frisch}, $\epsilon \sim u_{rms}^3/L$ where $u_{rms}^2$ is 
the mean square velocity. Since $u_{rms} \sim F$, we obtain that $F$ and $\epsilon$ are related as $\epsilon \sim F^3$.
By rescaling of  space, we
can always put $L=1$.  We fix $\epsilon=0.04$ and $\nu= 10^{-6}$, corresponding to an equivalent
$Re = u_{rms} L/\nu  \sim 3  \times 10^{5}$.
Most of our numerical results are
independent of $Re$ when $Re$ is large enough. In the limit $Re \rightarrow \infty$,  the
statistical properties of eq. (\ref{1d}) depend on 
the remaining free parameters, $D$, $\mu$ and $F$. 
For future reference, we compare the characteristic time scales for this simple model of homogeneous isotropic turbulence with the local doubling times of microorganisms described by Eqs. (1).     
Upon assuming the usual Kolmogorov scaling picture, we expect fluid mixing time scales $t$ in the range $\nu^{1/2}/\epsilon^{1/2} < t < L^{2/3}/\epsilon^{1/3}$,
or $0.01<t<3.0$, for typical parameter values of the shell model given above.    
On the other hand, the characteristic doubling time $t_2$
of, say, bacteria, in our model is $t_2 \sim 1/\mu$. 
Our simulations typically take $\mu=1$
so that $0.2 \le t_2 \le 1.0$,
implying cell division times somewhere in the {\it middle}  of the Kolmogorov range.  Microorganisms that grow {\it rapidly} compared to a range of turbulent mixing times out to the Kolmogorov outer scale, as is the case here, are crucial to the interesting effects we find when $\mu>0$. Bacteria or yeast, often mechanically shaken at frequencies of order 1Hz in a test tube in standard laboratory protocols, 
have cell division times of 20-90 minutes, and do not satisfy this criterion.   However, conditions that approximately match our simulations can be found for, say, bacterioplankton
 in the upper layer of the ocean, where large eddy turnover times do exceed microorganism doubling times \cite{review}, \cite{robinson}.
\begin{figure}[h]
  \begin{center}
    \includegraphics[width=0.5\textwidth]{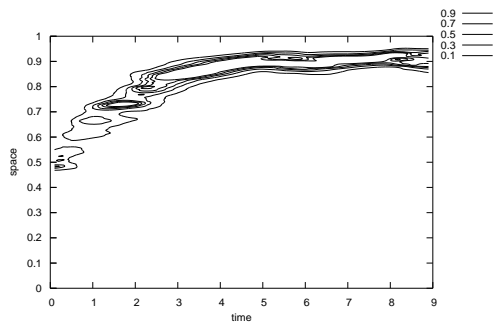}
  \end{center}
  \caption{ Same parameters and initial condition  as in Fig. (\ref{figure1}) for equation (\ref{1d}) with a "strong turbulent"  flow $u$ advecting  $c(x,t)$.
  }
  \label{figure2}
\end{figure}

\begin{figure}[h]
  \begin{center}
    \includegraphics[width=0.5\textwidth]{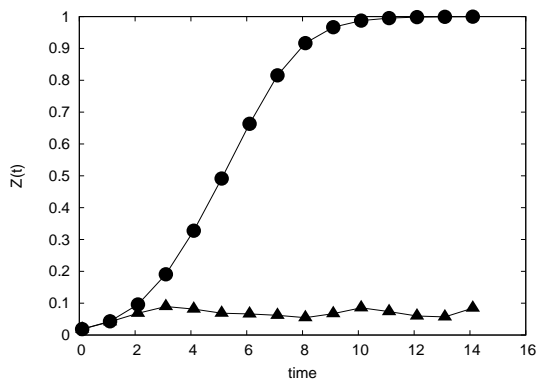}
  \end{center}
  \caption{ The behavior in time of the total "mass" $Z(t) \equiv \int dx c(x,t)$: circles show the function $Z$ for the case of Fig. (\ref{figure1}), i.e. a Fisher wave
with no turbulence; triangles show $Z$ for the case of Fig. (\ref{figure2}) when a strong turbulent flows is advecting $c(x,t)$
  }
  \label{figure3}
\end{figure}

In agreement with our previous theoretical analysis, in figure (\ref{fig2}) we show the numerical solutions of  Eq. (\ref{1d}) for a relatively "strong" turbulent flow.
A striking result is displayed:  we see no trace of a propagating front: instead, a well-localized pattern of $c(x,t)$ forms and stays more or less in a stationary
position. For us, Fig. (\ref{figure2}) shows a counter intuitive result. One naive expectation might be  that 
turbulence enhances mixing. The mixing effect due to turbulence is usually parametrized in the literature \cite{frisch}
by assuming an effective (eddy)  diffusion coefficient $D_{eff} \gg D$.
As a consequence, one  naive guess for Eq. (\ref{1d}) is that the spreading of an initial population is qualitatively similar to the travelling
Fisher wave with a more diffuse interface of width $\sqrt{D_{eff}/\mu}$
As we have seen,  this naive prediction is wrong for strong enough turbulence: the solution of
equation (\ref{1d}) shows  remarkable localized features which are preserved on time scales longer than the characteristic growth time $1/\mu$ or even the Fisher wave propagation time
 $L/v_F$. 
An important consequence of the localization effect is that the global "mass" (of growing microorganisms, say) ,  $Z \equiv \int dx c(x,t)$,  behaves differently with and without turbulence.
In Fig. (\ref{figure3}), we show $Z(t)$:
the curve with  circles refers to the conditions shown in Fig. (\ref{figure1})),
while the curve with triangles to Fig. (\ref{figure2}).  

The behavior of $Z$ for the Fisher equation without turbulence is a familiar {\it S}-shaped curve that reaches
the maximum $Z=1$ on a time scale $L/v_F$. On the other hand,  the effect of turbulence (because of
localization)   on the Fisher equation dynamics reduces significantly $Z$ almost by one order of magnitude

With biological applications in mind, it is important to determine conditions such that the spatial distribution of microbial organisms and the carrying capacity of the medium
are significantly altered by convective turbulence.
Within the framework of the Fisher equation, 
localization effect has been studied for a constant convection velocity and quenched time-independent spatial dependence in the
growth rate $\mu$  \cite{nelson1}, \cite{nelson2},  \cite{nelson3}, \cite{neicu}. In our case, localization,
when it happens, is  a time-dependent feature and depends on the statistical properties of the compressible turbulent flows. 
It is worth noting that the localized "boom and bust" population
cycles studied here may significantly effect 
"gene surfing"  \cite{genesurf} at the edge of a growing population, i.e. by changing the probability of gene mutation and fixation in the population.

\begin{figure}[h]
  \begin{center}
    \includegraphics[width=0.5\textwidth]{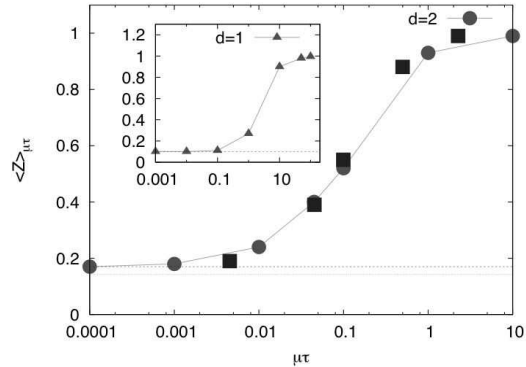}
  \end{center}
  \caption{ Behavior of the carrying capacity $\langle Z
    \rangle_\mu $ as a function of $\mu \tau$ from $128^2$ ( dots)
    and $512^2$ (squares) numerical simulations with $Sc=1$.
    Note that for $\mu\tau \rightarrow 0.001$, the carrying capacity
    approaches the limit $1/\langle P^2 \rangle$ (dotted line)
    predicted by Eq. (\ref{ZP2}). In the inset we show similar
    results for one dimensional compressible turbulent flows.}
  \label{fig4}
\end{figure}

One    prediction of eq.s (\ref{prima}) and (\ref{due}) is that the limit $\mu \rightarrow 0$ should be singular. More precisely, 
the quantity $\langle Z(t) \rangle $ must be equal to $1$ for $\mu=0$, while our predictions based
on (\ref{prima}) and (\ref{due}) imply that $\langle Z(t) \rangle < 1 $ for $\mu \ne 0$ because of "quasi localization" of the solutions. 
In the insert of figure (\ref{fig4}) we show the time averaged 
 $\langle Z(t) \rangle$,  computed for different values of $\mu$ for $F=0.5$. For large $\mu$, $<Z> \rightarrow 1$, as predicted by our
phenomenological approach, while in the limit $\mu \rightarrow 0$ the values of $<Z>$ converges to $0.1$. To predict the limit $\mu \rightarrow 0$  we can assume that 
 $c_{\mu}(x,t)$ for small enough
$\mu$ can be obtained by the knowledge of the solution $c_0(x,t)$ at $\mu=0$ by the relation
\begin{equation}
c_{\mu}(x,t) =^s   Z_{\mu}c_0(x,t)
\label{relation}
\end{equation}
where  in the above equation the symbol $=$ means "in the statistical sense"  and $ Z_{\mu} = \langle c_{\mu} \rangle_x $ (the subscript $x$ indicates average on space).
Since the solution $c_{\mu}(x,t)$ satisfies the constrain $\langle c_{\mu} \rangle_x - \langle c_{\mu}^2 \rangle_x = 0$ for any $\mu$, we obtain:
\begin{equation}
\label{zmu}
Z_{\mu} - Z_{\mu}^2 \langle c_0^2 \rangle_x =0 \ \ \rightarrow  \ \ Z_{\mu} =  \frac{1}{\langle c_0^2 \rangle_x}
\end{equation}
Once again we remark that eq. (\ref{zmu}) should be interpreted in a statistical sense, i.e. the time average of $Z_{\mu}$ should be equal for small $\mu$ to the time
average of $\langle c_0^2 \rangle_x^{-1}$. In the insert of figure (\ref{fig4}) the blue dotted line corresponds
to the time average of  $\langle c_0^2 \rangle_x^{-1}$: equation (\ref{zmu}) is clearly confirmed by our numerical findings.  As we shall see in the next section, the same argument can be applied for two dimensional
compressible flow.

\section{Fisher equation in two dimensional compressible flows}

As discussed in the previous section, an advecting compressible turbulent flow 
leads to highly non-trivial dynamics for the Fisher equation. Although previous results 
were obtained only in one dimension using a synthetic advecting flow from a 
shell mdel of turbulence, two striking effects were observed: the 
concentration field $c({\bm  x},t)$ is strongly localized near transient 
but long-lived sinks of the turbulent flows for small enough growth rate 
$\mu$; in the same limit, the space-time average concentration (denoted in 
the following as carrying capacity) becomes much smaller than its maximum 
value $1$.  Here, we  present numerical results aimed at
understanding the behavior of the Fisher'equation  for two dimensional
compressible turbulent flows and extending our previous results 
to more realistic two dimensional turbulent
flows.  Our model consists in assuming that the microorganism
concentration field $c({\bm x},t)$, whose dynamics is described by the equation
\begin{equation}
\label{eq:fish}
\partial_t c + {\bf \nabla} \cdot ({\bf u }c) = D \nabla^2 c+ \mu c(1-c)
\end{equation}
We assume that the population  is constrained on a planar surface of constant height
in a three dimensional fully developed turbulent flow with periodic
boundary conditions. Such a system could be a rough approximation to
microorganisms that actively control their bouyancy to mantain a fixed
depth below the surface of a turbulent fluid. As a consequence of this
choice, the flow field in the two dimensional slice becomes
compressible \cite{boffetta}.  We consider here a turbulent advecting
field ${\bf u}({\bf x},t)$ described by the Navier-Stokes equations,
and nondimensionalize time by the Kolmogorov time-scale
$\tau\equiv(\nu/\epsilon)^{1/2}$ and space by the Kolmogorov
length-scale $\eta\equiv(\nu^3/\epsilon)^{1/4}$. The non-dimensional
numbers charecterizing the evolution of the scalar field $C({\bf
  x},t)$ are then the Schmidt number $Sc=\nu/D$ and the
non-dimensional time $\mu\tau$.  A particularly interesting regime
arises when the doubling time $\mu^{-1}$ is somewhere in the middle of
the range of eddy turnover times that characterize the
turbulence. Although the underlying turbulent energy cascade is
somewhat different \cite{mck09}, this situation arises for oceanic
plankton, who double in $\sim 12$ hours, in a medium with eddy
turnover times varying from minutes to months \cite{mar03}.

\begin{figure}[h]
  \begin{center}
    \includegraphics[width=0.5\textwidth]{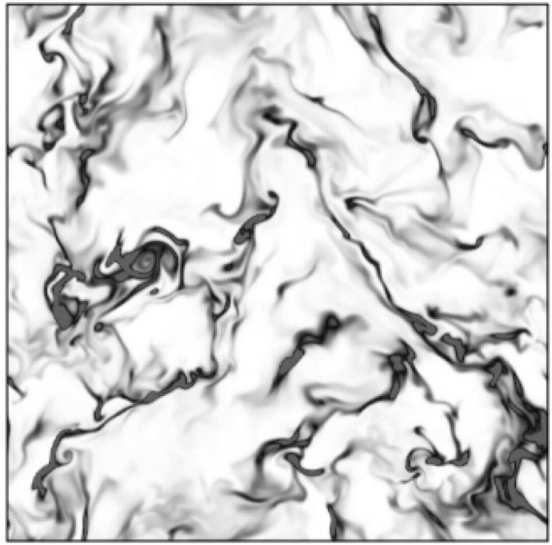}
  \end{center}
  \caption{ Plot of the concentration
      $c(x,y,t)$. The white color indicate regions with low
      concentration while regions of high concentration 
      are denoted by black.
  }
  \label{fig5}
\end{figure}

We conducted a three dimensional direct numerical simulation (DNS) of
homogeneous, isotropic turbulence at two different resolutions
($128^3$ and $512^3$ collocation points) {in a cubic box of length
  $L=2\pi$}.  The Taylor microscale Reynolds number \cite{frisch} for
the full 3D simulation was $Re_{\lambda}=75$ and $180$, respectively,
the viscosities were $\nu=0.01$ and $\nu=2.05 \cdot 10^{-3}$, the
total energy dissipation rate was around $\epsilon \simeq 1$ in both
cases. For the analysis of the Fisher equation we focused only on the
time evolution of a particular 2D slab taken out of the full three
dimensional velocity field and evolved a concentration field $c({\bm
  x},t)$ constrained to lie on this plane only. A typical plot of the
$2d$ concentration field, along with the corresponding velocity
divergence field (taken at time $t=86$, $Re_{\lambda}=180$) in the
plane is shown in Fig.~\ref{fig5} ($Sc=5.12$): the concentration $c(x,y,t)$ is highly peaked in small areas,
resembling one dimensional filaments. When the microorganisms grow
faster than the turnover times of a significant fraction of the
turbulent eddies, $c({\bf x},t)$ grows in a quasi-static compressible
velocity field, and accumulates near sinks and along slowly
contracting eigendirections, leading to filaments.  { The geometry
  of the concentration field suggests that $c(\bf x,t)$ is different
  from zero on a set of fractal dimension $d_F$ much smaller than
  $2$. A box counting analysis of the fractal dimension of $c(\bf
  x,t)$ supports this view and provides evidence that $d_F = 1. \pm
  0.15$.}

Note that for $\mu=0$, Eq. (\ref{eq:fish}) reduces to the
Fokker-Planck equation describing the probability distribution
$P(x,y,t)$ to find a Lagrangian particle subject to a force field 
${\bf u}({\bf x},t)$ at $x,y$ at time $t$:
\begin{equation}
  \label{fokkerplanck}
  {\frac{\partial P}{\partial t}} +  {\bm \nabla} \cdot ({\bm u} P) = D \nabla^2 P
\end{equation}
The statistical properties of $P$ have been studied in several works
(e.g. \cite{bec03} and \cite{massimo}) and it is known that for
compressible turbulence $P({\bf x},t)$ exhibits a nontrivial multifractal 
scaling. Upon multiplying eqn. (\ref{fokkerplanck}) by $P$ and integrating 
in space we obtain: $ \frac{1}{2} \partial_t \left\langle P^2\right\rangle_s +
\frac{1}{2} \left\langle P^2 (\nabla \cdot {\bf u}) \right\rangle_s =
- D \left\langle (\nabla P)^2 \right\rangle_s$ where $\left\langle
  \dots \right\rangle_s$ denotes a spatial integration. In the statistically 
stationary regime, the above equation reduces to:
\begin{equation}
\label{steadyP}
\frac{1}{2}  \langle  P^2 (\nabla \cdot {\bf u}) \rangle = - D \langle (\nabla P)^2 \rangle,
\end{equation} 
where now $\langle \dots\rangle$ stands for space and time
average. Eq. (\ref{steadyP}) shows that for $\nabla \cdot {\bf u}=0$
the only possible solution is $P={\mbox const}$. However, compressibility 
leads to nontrivial dynamics such that $P^2$ and $\nabla \cdot {\bf u}$ are 
anticorrelated. We measure the degree of compressibility by the factor 
$\kappa \equiv { \langle (\nabla \cdot {\bf u})^2 \rangle}/{\langle (\nabla
  {\bf u})^2 \rangle}$,  and estimate the l.h.s. of
Eq. (\ref{steadyP}) by assuming $ \langle P^2 (\nabla \cdot {\bf u})
\rangle = - A_1 \langle P^2 \rangle \langle (\nabla \cdot {\bf u} )^2
\rangle^{1/2}$, where we used the so called one point closure for
turbulent flows \cite{frisch} and $A_1$ is expected to be order unity.  We 
estimate the r.h.s of Eq. (\ref{steadyP}) by assuming:
\begin{equation}
\label{xi}
\langle (\nabla P)^2 \rangle = A_2 \frac{\langle P^2 \rangle}{\xi^2}
\end{equation}
where we define $\xi$ the ``quasi-localization'' length of $P$, which
is expected to be of the same order of the width of the narrow
filaments in Fig. \ref{fig5}. Finally we set $\langle (\nabla {\bf
  u} )^2\rangle = \epsilon/ \nu$ where $\epsilon$ is the mean rate of
energy dissipation and $\nu$ is the viscosity.  On putting everything
together we find a localization length given by:
\begin{equation}
\label{9}
\xi^2  =  \frac{2 A_2 D \sqrt{\nu} }{A_1 \sqrt{\kappa \epsilon}}.
\end{equation}
One important quantity -from the biological point of view- is the
carrying capacity, 
{
\begin{equation}
  Z(t) = \frac{1}{L^2}\int dx dy c({\bm x};t), 
\label{eq:carc}
\end{equation}
}
and in particular its time average in the statistical steady state with growth 
rate $\mu$, $\langle Z(t) \rangle_{\mu}$. We are interested to understand how 
$\langle Z \rangle_{\mu}$ behaves as a function of $\mu$, in the two 
important limits $\mu\rightarrow \infty$ and $ \mu \rightarrow 0$.
In the limit $\mu \rightarrow \infty$, we expect the carrying
capacity attains its maximum value $\langle Z \rangle_{\mu \rightarrow
  \infty} = 1$, because when the characteristic time $1/\mu$ becomes 
much smaller than the Kolmogorov dissipation time 
$\tau_{\eta} \equiv (\nu/\varepsilon)^{1/2}$, the effect of the velocity field 
is a relatively small perturbation on the rapid growth of the microorganisms. 
Indeed, consider a perturbation expansion of $c({\bf x},t)$ in
terms of $\delta=1/\mu$. On defining $c\equiv \displaystyle \sum_i \delta^ic_i({\bf x},t)$,
substituting in Eq.~(\ref{eq:fish}), assuming steady state, and
collecting the terms up to ${\cal O}(\delta^2)$ we find $c \approx
1 - \epsilon ({\bm \nabla} \cdot {\bm u}) + \epsilon^2 \{{\bm \nabla}
\cdot \left[{\bm u} ({\bm \nabla} \cdot {\bm u})\right] - D \nabla^2
(\nabla \cdot {\bm u})-({\bm \nabla} \cdot {\bm u})^2 \} + {\cal
  O}(\delta^3)$.  The above analysis  shows that in the limit
$\mu \to \infty$ the concentration field tends to become uniform with
the leading correction coming from the local compressibility. After 
substituting the expansion of $c$ in Eq.~(\ref{eq:carc}) one gets 
$Z \approx 1 - (\delta^2/L) \int ({\bm  \nabla} \cdot {\bm u})^2 {\rm {d \bm x}} + {\cal O}(\delta^3)$. Note that the leading correction to the carrying 
capacity is of order $\delta^2$, is consistent with the physical picture 
presented above.

By defining $\Gamma \equiv \langle ({\bm \nabla} \cdot {\bm u})^2
\rangle^{1/2}$ as the r.m.s value of the velocity divergence, we expect a
crossover in the behavior of $\langle Z \rangle_{\mu}$ for $\mu <
\Gamma$. In the limit $\mu \rightarrow 0$, following our discussion in the previous section, we expect that:
\begin{equation}
  \label{ZP2}
  \displaystyle \lim_{\mu \rightarrow 0 } \langle Z \rangle_{\mu} = \frac{1}{\langle P^2 \rangle}.
\end{equation}

We have  tested both
Eq.(\ref{ZP2}) and the limit $\mu\rightarrow \infty$ against our
numerical simulations.  In Fig. (\ref{fig4}) we show the behavior
of $\langle Z \rangle_{\mu}$ for the numerical simulations discussed
in this section. The horizontal line represents the value $ 1/\langle
P^2 \rangle$ obtained by solving Eq. (\ref{fokkerplanck}) for the same
velocity field and $\mu = 0$. For our numerical
simulations we estimate $\Gamma = 4.0$ and we observe, for $\mu >
\Gamma$ the carrying capacity $ \langle Z\rangle_{\mu}$ becomes close
to its maximum value $1$.  { The limit $\mu \rightarrow 0$
  requires some care. Let us define $\tau_b \equiv 1/\mu$ to be the time scale for the
  bacteria to grow.  The effect of turbulence is relevant for $\tau_b$
  longer than the Kolmogorov dissipation time scale $\tau_{\eta}$.  We
  also expect that $\tau_b$ must be smaller than the large scale
  correlation time $\tau_L \sim (L^2/\epsilon)^{1/3}$, which depends
  on the forcing mechanism driving the turbulent flows and the large
  scale $L$. Thus, the limit $\mu \rightarrow 0$ can be investigated
  either for $L \rightarrow \infty$ or by forcing the system with a
  constant energy input which slows down the large scale, as it is the
  case in our numerical simulations}

The limit $\mu \rightarrow 0$ can be investigated more accurately as
follows: according to known results on Lagrangian particles in
compressible turbulent flows, we know that $P$ should have a
multifractal structure in the inviscid limit $\nu \rightarrow 0$ \cite{14}. 
If our assumption leading to Eq. (\ref{ZP2}) is correct,  $c({\bf x},t)$  
must show multifractal behavior in the same limit with multifractal
exponents similar to those of $P$. For analytical results, see 
Ref. \cite{bec03} . Numerical evidence for the multifractal behavior of
Lagrangian tracers in compressible flows can be found in Ref. \cite{boffetta}.

We perform a multifractal analysis of the concentration field
$c(x,y,t)$ with $\mu>0$ by considering the average quantity
$\tilde{c}_{\mu}(r) \equiv \frac{1}{r^2} \int_{B(r)} c(x,y,t) dx dy $
where $B(r)$ is a square box of size $r$. Then the quantities $\langle
\tilde{c}_{\mu}(r)^p \rangle $ are expected to be scaling functions of
r, i.e.  $ \langle \tilde{c}_{\mu}(r)^p \rangle \sim r^{a(p)}$, where
$a(p)$ is a non linear function of $p$ with $a(2)=-0.47$, see \cite{18} for details.

\begin{figure}[h]
  \begin{center}
    \includegraphics[width=0.5\textwidth]{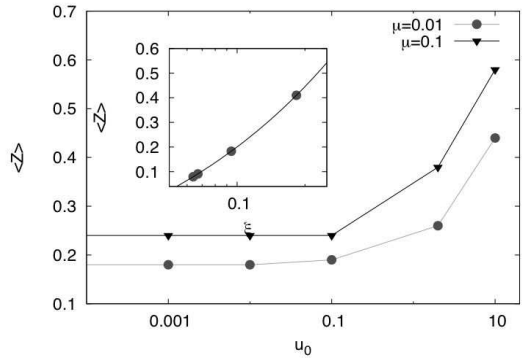}
  \end{center}
  \caption{ Main figure: plot of $ \langle Z \rangle$ as function of a
    super imposed external velocity $u_0$ for $\mu=0.01$ (bullets)
    and $\mu=0.1$ (triangles). Inset: log-log plot of $\langle Z
    \rangle$ as a function of the localization length $\xi$ defined in
    Eq. (\ref{xi}) for $u_0=0$.  The slope is consistent with the {
    prediction $<Z> \sim \xi^{-a(2)}$} discussed in the text. The numerical 
    simulations are done for $\mu=0.01$ and different values of $D$ from 
    $D=0.05$ to $D=0.001$. 
  }
  \label{fig6}
\end{figure}

{Our multifractal analysis allow us to investigate the possible
relation between the localization length $\xi$ defined in
Eq. (\ref{xi}) and the carrying capacity $\langle Z \rangle_{\mu}$.
The localization length $\xi$ can be considered as the smallest scale below which one should observe fluctuations
of $c(x,t)$. Thus we can expect that $\langle P^2 (x,t) \rangle \sim \xi^{a(2)}$. Using (\ref{ZP2}) we obtain 
$\langle Z \rangle \sim \xi^{-a(2)}$.
In the
inset of Figure (\ref{fig6}) we show $\langle Z \rangle$ as a
function of $\xi$ (obtained by using (\ref{xi}) for $\mu=0.01$ and
different values of the diffusivitiy $D$. According to Eq. (\ref{9}), 
reducing the diffusivity $D$ will shrink the localization length $\xi$ 
and hence $\langle Z \rangle_{\mu}$. From Figure (\ref{fig6}) a 
clear power law behavior is observed with a scaling exponent $0.46$ very
close to the predicted behavior $-a(2) = 0.47$.}

Finally, we discuss bacterial populations subject to both
turbulence and uniform drift because of, e.g., sedimentation under 
the action of gravity field. In this case, we can
decompose the velocity field into zero mean turbulence fluctuations plus a
constant ``wind'' velocity $u_0$. In presence of a mean drift velocity 
Eq.~\ref{eq:fish} becomes:
\begin{equation}
  {\frac{\partial c}{\partial t}} + {\bm \nabla} \cdot [({\bm u} + u_0 \hat{e_x})c] = D \nabla^2 c +  \mu c(1 -  c)
\end{equation}
where $\hat{e_x}$ is the unit vector along the $x$-direction. Note
that the mean drift  breaks the Galilean invariance as the concentration 
$c$ is advected by the wind, while turbulent fluctuations ${\bf u}$ remain 
fixed. In Fig.~\ref{fig6} we show the variation of carrying capacity 
versus $u_0$ for two different values of $\mu$ and fixed
diffusivity $D=0.015$. We find that for $u_0 \leq  u_{rms}$ ($u_{rms}$ is
the root-mean-square turbulent velocity) the carrying capacity $Z$ saturates to
a value equal to the value of $Z$ in absence of $u_0$ i.e., quasilocalization 
by compressible turbulence  dominate the dynamics. For $u_0>u_{rms}$ the 
drift velocity delocalizes the bacterial density thereby causing $Z \to 1$, in agreement with the
results discussed in \cite{1d}.

\section{Discrete population dynamics}

The population dynamics of a
single species expanding into new territory was first studied in the
pioneering works of Fisher, Kolmogorov, Petrovsky and Piscounov (FKPP)
\cite{fisher,2,3}. Later, Kimura and Weiss studied individual-based
counterparts of the FKPP equation \cite{6}, revealing the important
role of number fluctuations. In particular, stochasticity is
inevitable at a frontier, where the population size is small and the
discrete nature of the individuals becomes essential.  Depending on
the parameter values, fluctuations can produce radical changes with
respect to the deterministic predictions \cite{3,4}.  If $f(x,t)$ is
the population fraction of, say, a mutant species and $1-f(x,t)$ that
of the wild type, the stochastic FKPP equation reads in one dimension
\cite{8}:
\begin{equation}\label{fkpp}
\partial_t f(x,t) = D\partial^2_x f + s f(1-f) +\sqrt{D_g f(1-f)}\xi (x,t)
\end{equation}
where $D$ is the spatial diffusion constant, $D_g$ is the genetic
diffusion constant (inversely proportional to the local population
size), $s$ is the genetic advantage of the mutant and
$\xi=\xi(x,t)$ is a Gaussian noise, delta-correlated in time and space
that must be interpreted using Ito calculus \cite{8}. In the neutral
case ($s=0$), number fluctuations induce a striking effect in the
population dynamics, namely segregation of the two species. One can
show that the dynamics of competing species in 1D can be characterized
by the dynamics of boundaries between the $f=0$ and $f=1$ states of
Eq. \ref{fkpp}, which perform a random walk. This effect is
theoretically predicted by Eq.(\ref{fkpp}) and confirmed
experimentally in the linear inoculation experiments on neutral
variants of fluorescently labelled bacteria illustrated in
Fig. (\ref{fig1}a) \cite{hall}.

We study the influence of advection on the dynamics of
two distinct populations consisting of discrete 'particles'. 
Due to competition and stochasticity,
interactions between two populations usually drive one of the two
populations to extinction. The average time of this event (the {\it
fixation time}) is a quantity of
great biological interest since it determines the amount of genetic
and ecological diversity that the system can sustain. Studying
competition in a hydrodynamics context, where both a compressible
velocity field and stochasticity due to finite population sizes are
present, calls for a nontrivial generalization of
Eq. (\ref{fkpp}). One complication is that, because of
compressibility, the sum of the concentrations of the two species is
no longer invariant during the dynamics. 
\begin{figure}[h]
  \begin{center}
    \includegraphics[width=0.5\textwidth]{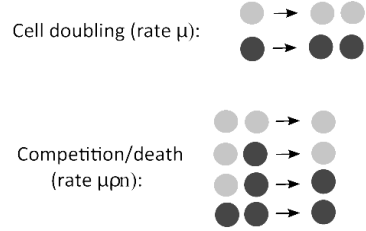}
  \end{center}
  \caption{The possible six birth and death processes in the particle model
consisting of two species, A (red) and B (green). 
  }
  \label{fig7}
\end{figure}
\begin{figure}[htb]
\begin{center}
\includegraphics[width=8.8cm]{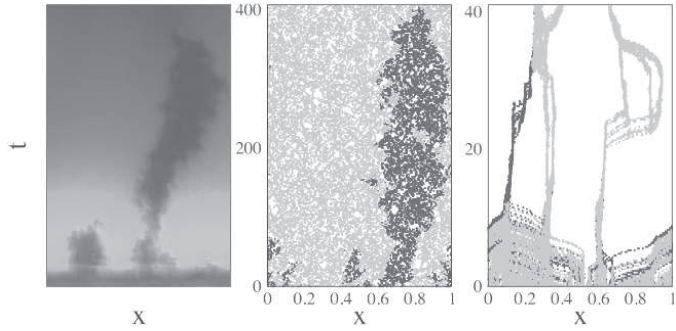}
\caption{(a) Experimental range expansion of the two neutra E. coli
  strains used in Ref. \cite{hall}, but run about one day longer
  (D. Nelson, unpublished). The black bar at the bottom is due a small
  crack in the agar substrate.  (b) Space-time plot of the off-lattice
  particle model with no advecting velocity field. A realization
  characterized by a pattern similar to the experimental one has been
  selected for illustrative purposes. (c) Particle model with a
  compressible turbulent velocity field. Simulations are run until
  fixation (disappearance of one of the two species); note the reduced
  carrying capacity and the much faster fixation time in
  (c). Parameters: $N=10^3$, $D= 10^{-4}$, $\mu=1$. Parameters of the
  shell model are as in \cite{17}. \label{fig1} }
\end{center}
\end{figure}

We have overcome these problems through a
off-lattice particle model designed to explore how compressible
velocity fields affect biological competition. Let us consider two
different organisms, $A$ and $B$, which advect and diffuse in space
while undergoing duplication (i.e. cell division) and
density-dependent annihilation (death), see Fig. \ref{fig7}. 
Specifically, we implement the
following stochastic reactions: each particle of species $i=A,B$
duplicates with rate $\mu_i$ and annihilates with a rate $\bar{\mu}_i
\widehat{n}_i$, where $\widehat{n}_i$ is the number of neighboring
particles (of both types) in an interaction range $\delta$. Let $N$ be
the total number of organisms that can be accomodated in the unit
interval with total density $c_A+c_B=1$.  To reduce the number of
parameters, we fix $\delta=1/N$ as the average particle spacing in the
absence of flow. Further, we set $\bar{\mu}_A=\bar{\mu}_B=\mu_B=\mu$,
but take $\mu_A=\mu(1+s)$ to allow for a selective advantage (faster
reproduction rate) of species $A$. We will start by analyzing in depth
the neutral case $s=0$ and consider the effect of $s>0$ in the end of
the Letter.  In one dimension and with these choices of parameters,
our macroscopic coupled equations for the densities $c_A(x,t)$ and
$c_B(x,t)$ of individuals of type $A$ and $B$ in an advecting field
$v(x,t)$ read
\begin{eqnarray}\label{model}
\partial_t c_A\!\!&=\!\!&\!
-\partial_x(vc_A)\!+\!D\partial^2_xc_A\!+\!\mu c_A(1\!+\!s\!-\!c_A\!-\!c_B)\!+\!
\sigma_A\xi
\nonumber\\
\partial_t c_B\!\!&=\!\!&\!
-\partial_x(vc_B)\!+\!D\partial^2_xc_B\!+\!\mu c_B(1\!-\!c_A\!-\!c_B)\!+\!
\sigma_B\xi'\ 
\end{eqnarray}
with $\sigma_A=\sqrt{\mu c_A(1+s+c_A+c_B)/N}$ and $\sigma_B=\sqrt{\mu
  c_B(1+c_A+c_B)/N}$. $\xi(x,t)$ and $\xi'(x,t)$ are independent
delta-correlated noise sources with an Ito-calculus interpretation as
in Eq. (\ref{fkpp}).
\begin{figure}[htb]
\begin{center}
\includegraphics[width=8.4cm]{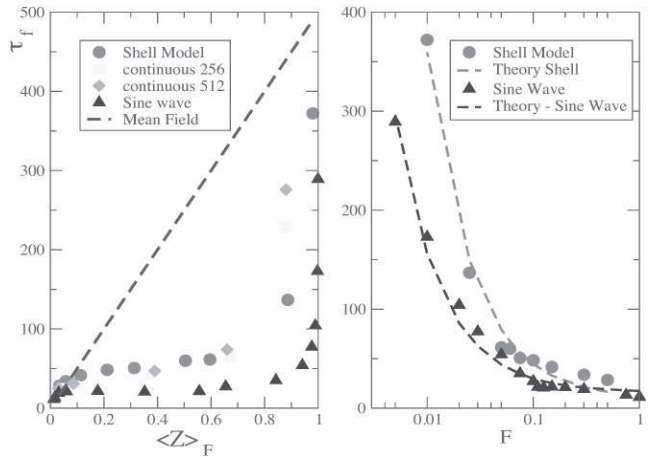}
\caption{Average fixation time $\tau_f$ for neutral competitions in
  compressible turbulence and sine wave advection, as a function of
  (left) the reduced carrying capacity $\langle Z\rangle_F$ and
  (right) forcing intensity $F$ (small $\langle Z\rangle_F$ in the
  left panel corresponds to large forcing in the right panel).
  (left) Red circles and blue triangles are particle
  simulations. Other symbols denote simulations of the continuum
  equations with different resolutions on the unit interval. The black
  dashed line is the mean field prediction, $\tau_f=N\langle
  Z\rangle_F/2$. In (right), only particle simulations are shown and
  dashed lines are the theoretical prediction $\tau_f=\tau_0+c/F$
  based on boundary domains, with fitted parameters $\tau_0=9.5$,
  $c=3.5$ in the case of the shell model and $\tau_0=16$, $c=1.4$ in
  the case of the sine wave.\label{fig2}}
\end{center}
\end{figure}

Simulations of the particle model corresponding to (\ref{model}) with
$v=s=0$ result in a dynamics similar to the one observed in
experiments, as shown in Fig.(\ref{fig1}b). In this simple limit, our
model can be considered as a grand canonical generalization of Eq
(\ref{fkpp}), where the total density of individuals $c_A + c_B$ is
now allowed to fluctuate around an average value $1$. 
We fix the following
parameters: $N=10^3$, $D= 10^{-4}$, $\mu=1$ and $L=1$ where $L$ is a
one dimensional domain endowed with periodic boundary conditions. With
these parameters, the fixation time $\tau_f$ would be $\sim 10^4$ for
the one dimensional FKKP equation, and $\sim 10^3$ for the well-mixed
case.

Introducing a compressible velocity field $v(x,t)$, via the shell
model \ref{sabra} as shown in
Fig.(\ref{fig1}c), leads to radically different dynamics.  Individuals
tend to concentrate at long-lived sinks in the velocity
field. Further, extinction is enhanced and the total number of
individuals $n(t)$ present at time $t$ is on average smaller than $N$.

In order to study how a velocity field changes $ \tau_f$, we first
analyze two different velocity fields: The first is a velocity field
$v(x,t)$ generated by a shell model (Eqs. \ref{sabra}) of compressible turbulence
\cite{17}, reproducing the power spectrum of high Reynolds number
turbulence with forcing intensity $F$. The second is a
static sine wave, $v(x)=F \sin(2\pi x)$, representing a simpler case
in which only one Fourier mode is present, and thus a single sink, in
the advecting field. In both cases, periodic boundary conditions on
the unit interval are implemented.

Fig.(\ref{fig2}) shows the average fixation time $\tau_f$ for $s=0$ in
the first two cases, while varying the intensity $F$ of advection.  In
the left panel, we plot the fixation times as a function of the
time-averaged reduced carrying capacity $\langle Z \rangle_F$, where
$Z(t)=n(t)/N$ is the carrying capacity reduction, i.e. the ratio
between the actual number of particles and the average number of
particles $N$ observed in absence of the velocity field.  Plotting
vs. $\langle Z \rangle_F$ allows comparisons with the mean field
prediction, $\tau_f=2N \langle Z\rangle_F/\mu$, valid for well mixed
systems (black dashed line) \cite{8}.  For the shell model, we include
simulations of the macroscopic equations (\ref{model}) with different
resolutions ($256$ and $512$ lattice sites on the unit interval),
obtaining always similar results for $\tau_f$ vs. $\langle Z
\rangle_F$.

In all cases, the presence of a spatially varying velocity field leads
to a dramatic reduction of $\tau_f$, compared to mean field
theory. The fixation time drops abruptly as soon as $\langle Z \rangle
<1$, even for very small $F$.

\bigskip

{\bf Acknowledgment}

We acknowledge computational support from CASPUR (Roma, Italy uner HPC 
Grant 2009 N. 310), from CINECA (Bologna, Italy) and SARA (Amsterdam, The
Netherlands). Support for D.R.N. was provided in part by the National 
Science Foundation through Grant DMR-0654191 and by the Harvard Materials 
Research Science and Engineering Center through NSF Grant DMR-0820484. 
Data from this study are publicly available in unprocessed raw format 
from the iCFDdatabase (http://cfd.cineca.it). M.H.J. was supported by
Danish National Research Foundation through "Center for Models of Life".


\begin{thebibliography}{10}

\bibitem{3} W. van Saarloos,
   Phys. Rep. {\bf 386}, 29-222 (2003).

\bibitem{4} O. Hallatschek and K.  Korolev, Phys. Rev. Lett. {\bf
    103}, 108103 (2009), and references therein.

\bibitem{fisher}
R. Fisher, Ann. Eugenics {\bf 7},  335  (1937); A. Kolmogorov, 
I. Petrovsky and N. Psicounoff, Moscow, Univ. Bull. Math, {\bf 1}, 1, (1937).

\bibitem{6} M. Kimura and G. H. Weiss, Genetics {\bf 49}, 561-576
   (1964); J. F. Crow and M. Kimura {\em An Introduction to Population
     Genetics}, Blackburn Press, Caldwell, NJ (2009).


\bibitem{8} For a recent review, see K. Korolev et al.
    Rev. Mod. Phys. {\bf 820}, 1691-1718 (2010).


\bibitem{11} B. A. Whitton and M. Potts {\em The Ecology of
      Cyanobacteria: Their Diversity in Time and Space} eds.  Kluwer,
    Dordrecht, Netherland

\bibitem{12} F. D'Ovidio et al.,
   Proc. Natl. Acad. Sci. {\bf 107},
  18366-18370 (2010).
ds (2000).


\bibitem{13} W. J. McKiver and Z.  Neufeld,
  Phys. Rev. E {\bf 79}, 061902 1-8 (2009).





\bibitem{dissipation} F. Peters, C. Marraese, Marine Ecology Progress Series, {\bf 205}, 291, (2000)

\bibitem{Toschi} F. Toschi, E. Bodenschatz, Annual Rev. Fluid. Mech., {\bf 41}, 375, (2008)

\bibitem{14} J. Bec, Phys. Fluids {\bf 15}, L81-L84 (2003).


\bibitem{Gyrotaxis1} W. M. Durham, E. Ciment, R. Stocker, Phys. Rev. Lett., {\bf 106}, 238102, (2011)

\bibitem{Gyrotaxis2} C. Torney, Z. Neufeld, Phys. Rev. Lett., {\bf 99}, 078101,(2007)


\bibitem{klein} P. Klein and G. Lapeyre, Annual Review of Marine Science, {\bf 1}, 357, (2009)

\bibitem{Mizobota} K. Mizobota, Saitoh SI, Shiomoto A., Miyamura T, Shiga N, et. al , Prog. Oceanogr., {\bf 55} , 65, (2002)

\bibitem{15}  A. P. Martin,
Progr. Ocean. {\bf 57}, 125-174  (2003).

\bibitem{SQG1} I. M. Held, R. T. Pierrehumber, S.T. Garner, K.L. Swanson, J. Fluid. Mech., {\bf 282}, 1, (1995)

\bibitem{SQG2} X. Capet, P. Klein, B.L. Hua, G. Lapeyre, J. C. McWilliams, J. Fluid. Mech., {\bf 604}, 165, (2008)


\bibitem{3docean} P. Klein, B. L. Hua, G. Lapeyre, X. Capet, S. Le Gentil, H. Sasaki, J. Phys. Oceanogr. {\bf 38}, 1748, (2008)

\bibitem{fronts1} L. Thomas, A. Tandon, A. Mahadevan, J. Geophys. Res., {\bf 177}, 17, (2008)

\bibitem{17} R. Benzi, D. R. Nelson
  Physica D {\bf 238} 2003-2015  (2009).

\bibitem{18} P. Perlekar, R. Benzi, D.R. Nelson, 
Phys. Rev. Lett. {\bf 105}, 144501 (2010). 

\bibitem{18b} S. Pigolotti, R. Benzi, M.H. Jensen, D. R. Nelson, Phys. Rev. Lett. submitted.


\bibitem{2} A. Kolmogorov, N. Petrovsky, and N. Piscounov, Moscow
  Univ. Math. Bull. {\bf 1}, 1-25 (1937).


 \bibitem{vulpiani2} S. Berti, D. Vergni, A. Vulpiani
 Europhys. Lett. 83, 54003 (2008)

\bibitem{biferale} L. Biferale, Annu., 2003, Rev. Fluid Mech. 35, 441.
\bibitem{frisch}
U. Frisch, {\em Turbulence the legacy of A.N. Kolmogorov} (Cambridge University
  Press, Cambridge, 1996).



\bibitem{review} T. Tel et. al., Chemical and Biological Activity in Open Flows: A Dynamical Systems Approach, Phys. Reports, 413, 91, 2005

\bibitem{robinson}A. R. Robinson, Proc. R. Soc. Lond. A453, 2295 (1997);   A455, 1813 (1999)


\bibitem{nelson1} D. R. Nelson, and N. M. Shnerb. 1998. Non-hermitian localization and
population biology. Phys. Rev. E. 58:1383.

\bibitem{nelson2}K.A.  Dahmen, D. R. Nelson, and N. M. Shnerb. 2000. Life and death near
a windy oasis. J. Math. Biol. 41:1-23.

\bibitem{nelson3} N. M. Shnerb,  2001, Extinction of a bacterial colony under forced
convection in pie geometry. Phys. Rev. E 63:011906, and references therein.

\bibitem{neicu} T. Neicu,  A. Pradhan, D. A. Larochelle, and A. Kudrolli. 2000. Extinction
transition in bacterial colonies under forced convection. Phys. Rev. E. 62:1059 - 1062.

\bibitem{genesurf} O. Hallatschek and D. R. Nelson, Theor. Popul. Biology, 73,1, 158,  2007.

\bibitem{16}J. R. Cressman et al.,
  Europhys. Lett. {\bf 66}, 219-225 (2004); G. Boffetta et al.,
  Phys. Rev. Lett. {\bf  93}, 134501  (2004).

\bibitem{Wakika}
J. ichi Wakita {\it et~al.}, J. Phys. Soc. Jpn. {\bf 63},  1205  (1994).

\bibitem{1d}
R. Benzi and D. Nelson, Physica D {\bf 238},  2003  (2009).

\bibitem{boffetta}
G. Boffetta, J. Davoudi, B. Eckhardt, and J. Schumacher, Phys. Rev. Lett. {\bf
  93},  134501  (2004).

\bibitem{bec03}
J. Bec, Phys. Fluids {\bf 15},  L81  (2003);  J. Bec, J. Fluid Mech., 
{\bf 528}, 255 (2005).

\bibitem{massimo}
G. Falkovich, K. Gawedzki, and M. Vergassola, Rev. Mod. Phys. {\bf 73},  914
  (2001).


\bibitem{hall} O. Hallatschek and D. Nelson, Proc. Natl. Acad. Sci
  {\bf 104}, 19926-19930 (2007).


\bibitem{mck09}
W. McKiver and Z. Neufeld, Phys. Rev. E {\bf 79},  061902  (2009).

\bibitem{mar03}
A. Martin, Prog. in Oceanography {\bf 57},  125  (2003).


\bibitem{kur00}
A. Kurganov and E. Tadmor, J. Comp. Phys. {\bf 160},  241  (2000).


\end{thebibliography}

\end{document}